\documentclass{statsoc}

\usepackage{amsmath,amssymb,dsfont}
\usepackage{natbib}
\usepackage{nameref,hyperref}

\usepackage{bm}
\usepackage[a4paper]{geometry}
\usepackage{graphicx}

\title[Intrinsic dimensionality of Covid-19 data]{On the intrinsic dimensionality of Covid-19 data: a global perspective}
\author[Varghese et al.]{Abhishek Varghese \textsuperscript{1,2}}
\author{Edgar Santos-Fernandez  \textsuperscript{1,2}}
\author{Francesco Denti  \textsuperscript{3}}
\author{Antonietta Mira  \textsuperscript{4,5}}
\author[Varghese et al.]{Kerrie Mengersen  \textsuperscript{1,2}}

\address{\textbf{1} School of Mathematical Sciences, Queensland University of Technology, Brisbane, Australia}
\address{\textbf{2} ARC Centre for Excellence in Mathematical and Statistical Frontiers (ACEMS), Brisbane, Australia. Funded by the Centre for Data Science (CDS) at Queensland University of Technology (QUT), Brisbane, Australia.}
\address{\textbf{3} Department of Statistics, Università Cattolica del Sacro Cuore, Milan, Italy.}
\address{\textbf{4} Institute of Computational Science, Università della Svizzera italiana, Lugano, Switzerland.}
\address{\textbf{5} Department of Science and High Technology, Università degli Studi dell’Insubria, Como, Italy.}

\email{varghese.abhishek@gmail.com}

\begin{document}
\section{Abstract}
This paper aims to develop a global perspective of the complexity of the relationship between the standardised per-capita growth rate of Covid-19 cases, deaths, and the OxCGRT Covid-19 Stringency Index, a measure describing a country's stringency of lockdown policies. To achieve our goal, we use a heterogeneous intrinsic dimension estimator implemented as a Bayesian mixture model, called Hidalgo. We identify that the Covid-19 dataset may project onto two low-dimensional manifolds without significant information loss. The low dimensionality suggests strong dependency among the standardised growth rates of cases and deaths per capita and the OxCGRT Covid-19 Stringency Index for a country over 2020-2021. Given the low dimensional structure, it may be feasible to model observable Covid-19 dynamics with few parameters. Importantly, we identify spatial autocorrelation in the intrinsic dimension distribution worldwide. Moreover, we highlight that high-income countries are more likely to lie on low-dimensional manifolds, likely arising from aging populations, comorbidities, and increased per capita mortality burden from Covid-19. Finally, we temporally stratify the dataset to examine the intrinsic dimension at a more granular level throughout the Covid-19 pandemic.
\newpage

\section{Introduction}
High-dimensional datasets are generally challenging for statistical inference and data analysis. Moreover, their analysis is made even more challenging by longitudinal measurements and temporal autocorrelation. Fortunately, these kinds of data have a high degree of redundancy typically and, therefore, may project onto low-dimensional manifolds without losing substantive information \citep{levina2005maximum, camastra_intrinsic_2016}. The dimensionality of these manifolds is called the \emph{intrinsic dimension} (ID) of the data, and it can provide important information about the properties of datasets.

Data science methods for high-dimensionality datasets have been utilised and explored in multiple contexts to aid decision-making and analysis during the Covid-19 pandemic. For example, citywide smart card travel data in Sydney, Australia, has been utilised to cluster passenger types along multiple mobility dimensions and develop intervention strategies for disease spread \citep{shoghri_identifying_2020}. Similarly, manifold learning techniques have been applied to cell-phone mobility data in the United States during the Covid-19 pandemic, distinguishing mobility trends in multiple geographic regions and demographics \citep{levin_cell_2020}. \cite{wisesty_comparison_2021} leverage dimensionality reduction techniques to cluster and analyse highly dimensional genome sequence data of Covid-19. Similarly,  \cite{hearn_higher-ed_2020} identify essential features in predicting the mode of instruction in American universities during the Covid-19 pandemic. 

Additionally, these statistical techniques have enabled decision-makers to parse the large body of communication transmitted online during the Covid-19 pandemic to glean new insights. \cite{ordun_exploratory_2020} employ Uniform Manifold Approximation and Projection and Latent Dirichlet Allocation to parse Twitter data during the Covid-19 pandemic and distinguish topics, identify trends and patterns in social network behaviours. \cite{doanvo_machine_2020} utilise similar techniques to analyse a large body of open access Covid-19 research studies and classify research output to identify existing knowledge gaps in research.

However, there has been little work done to explore the latent dynamics of the pandemic spread across continents and countries. \cite{sivakumar_complexity_2021} examine the temporal dynamics of Covid-19 daily cases and deaths in 40 countries, using a False Nearest Neighbour method to identify the relevant embedding dimension (ED) for each country. The authors recognise that new Covid-19 cases and deaths exhibit a low- to medium-level ED. However, it is essential to note that ED does not account for points in a dataset lying on low-dimensional manifolds. Thus, identifying the ID is generally more valuable as it accounts for inherent structures in the data and remains a more accurate representation of underlying structural complexity in a dataset \citep{Santos-Fernandez2021, eneva2002wekkem}. This research work will seek to bridge this gap and provide valuable information towards understanding the complexity and dimensionality of the Covid-19 pandemic in different countries and develop a deeper understanding of the spread of the pandemic.

This paper provides an application of the recent heterogeneous ID algorithm (Hidalgo). Hidalgo is a Bayesian mixture model capable of clustering the observations into groups characterised by similar IDs. The ID can be considered an indicator of the complexity of the data: the higher its value, the larger the number of relevant directions are required to represent the data points faithfully. More information about ID may be found in the next section, and more formal definitions of ID can be found in \cite{bishop_neural_1995} and \cite{camastra_intrinsic_2016}.

The vast majority of statistical methods assume and estimate a unique value for the ID. However, this assumption is often too strong for datasets containing information generated by intricated systems with complex dynamics, such as a global pandemic. Hidalgo extends this framework, allowing the presence of multiple manifolds characterised by different ID values in the same dataset. The Bayesian local ID estimator has been applied successfully to a diverse range of datasets for scenarios such as financial markets, neuroimaging, proteomics \citep{Allegra}, genomics \citep{denti_intrinsic_2022}, and high-resolution player tracking data \citep{Santos-Fernandez2021}. Here, we seek to organise the pandemic dynamics of different countries into groups with similar ID to help us unveil non-trivial patterns related to the dynamics of the Covid-19 pandemic. Finally, we temporally stratify the dataset to examine the ID at a more granular level throughout the Covid-19 pandemic. 

\section{Materials and methods}

\subsection{Likelihood-based intrinsic dimension estimation}
A large number of ID estimators are currently available in the literature. Likelihood-based estimators are particularly appealing because of their theoretical foundation and the immediate ability to provide estimates for uncertainty quantification. 

Recently, building on the work of \cite{levina2005maximum} and \cite{Comment}, \cite{facco2017estimating} introduced the `Two Nearest Neighbour' (TWO-NN) estimator, based on the following distributional result. Assume we observe $n$ units in a dataset of nominal dimension $D$ (intuitively, the number of recorded columns in a tall dataset), where the data lies on a manifold of smaller dimension $d$, the ID. In other words, some dimensions are irrelevant. Furthermore, we may consider the dataset as a configuration obtained from a Poisson point process over $\mathbb{R}^D$ characterised by a homogeneous intensity function $\rho$. In that case, the ratio of the distances between a given point and its second and first nearest neighbours (NN) is Pareto distributed with shape parameter $d$ and scale parameter identically equal to 1. Algebraically, denoting with $r_{i,j}$ the distance between the $i$-th point and its $j$-th NN, we have:
  \begin{equation}
\mu_{i} = \frac{r_{i,2}}{r_{i,1}} \sim Pareto(1,d), \quad \quad \mu_i \in \left(1,+\infty\right) \quad \quad i=1,\ldots,n.
\label{eq::firstresult}
\end{equation}
Although the theoretical derivation requires a uniform intensity of the point process, the result in Eq.~\eqref{eq::firstresult} is empirically valid as long as the homogeneity assumption holds up to the second NN for every point.

As previously mentioned, methods that return a unique ID value to describe the entire dataset can often be limiting and unrealistic since data may lie on multiple latent ID manifolds. To address this shortcoming, \cite{Allegra} extended the TWO-NN by partitioning the data in subgroups characterised by locally homogenous ID via a Bayesian mixture model \citep{gelman1995bayesian}. The ratios $\mu_{i,1,2}$ now arise from $L$ different Pareto distributions, obtaining:
\begin{equation}
f(\mu_{i}|\bm{d},\bm{\pi}) = \sum_{l=1}^{L}\pi_l \: d_l \mu_i^{-(d_l+1)}, \quad \quad i=1,\ldots,n,
\label{hid}
\end{equation}

\noindent where $\bm{\pi}=\left(\pi_1,\ldots,\pi_L\right)$ is the vector of mixture weights and $\bm{d}=\left(d_1,\ldots,d_L\right)$ is a vector containing the ID parameters.
The Bayesian model is completed with prior distribution specifications. In particular, the authors chose identically distributed and independent Gamma priors for each element of $\bm{d}$, with shape and rate parameters $a_d>0$ and $b_d>0$, respectively : $d_l\sim Gamma(a_{d},b_{d})\:\:\forall\:\:l$. Morevoer, a Dirichlet prior for the mixture weights $\bm{\pi}\sim Dirichlet(\alpha_1,\ldots, \alpha_L)$, where $\bm{\alpha}=(\alpha_1,\ldots, \alpha_L)$ is a vector of positive concentration parameters. Careful choice of $\bm{\alpha}$ allows the model to automatically select the necessary number of mixture components $L^*\leq L$. Within this context, the value $L$ in Eq.~\eqref{hid} is now interpreted as an upper bound on the number of populated clusters. As customary in Bayesian mixture models, we can augment the parameter space to enhance inference and ease posterior computation, adding the auxiliary parameters $c_i\in\{1,\ldots,L\}$, for $i=1,\ldots,n$.

These latent membership labels link each observation to a cluster. In other words, $c_i=l$ implies that the $i$-th unit corresponds to the $l$-th mixture component.
Unfortunately, even given this expansion to the model space, fitting the model presented in Eq.~\eqref{hid} is exceptionally challenging: the overlaying support of the Pareto distributions jeopardises the clustering assignment, which in turn prevents the derivation of reliable estimates of the ID. To deal with this issue, \cite{Allegra} alter the model by introducing a local homogeneity assumption, leading to the following modified likelihood:
\begin{equation}
\mathcal{L}\left(\mu_i,\mathcal{N}^{(q)}|\bm{d},\bm{c},\zeta\right) =  \: d_{c_i} \mu_i^{-(d_{c_i}+1)}\times \prod_{i=1}^{n}\frac{ \zeta^{\mathds{1}_{c_i\neq c_j}}(1-\zeta)^{\mathds{1}_{c_i=c_j}}}{\mathcal{Z}_i}, \quad \quad c_i|\bm{\pi} \sim Cat_{L}(\bm{\pi}),
\label{MODpara}
\end{equation}
where $Cat_{L}$ denotes a Categorical distribution over the set $\{1,\ldots,L\}$. A closed form expression for the posterior distribution is not available, so we rely on Markov Chain Monte Carlo (MCMC) techniques to simulate a posterior sample. The additional product term in the likelihood is needed to enforce that points close to each other are more likely to be part of the same latent manifold. More technical discussions of this model specification and the validity of the underlying hypothesis can be found in the Supplementary Material of \cite{Allegra} and \cite{denti_intrinsic_2022}. In the same references, one can find more details about the Gibbs sampler algorithm used for fitting the model and the post-processing tools adopted to deal with computational issues such as label-switching. In this paper, we apply the model defined by Eq.~\eqref{hid} and the corresponding Hidalgo algorithm developed by \cite{Allegra} to an assessment of global Covid-19 disease dynamics. More details are provided in the following subsections.

\subsection{Data Description}
This work utilises three datasets to explore the disease and spreading dynamics of Covid-19 in countries: Covid-19 new cases, deaths per million population (pmp) \cite{ritchie_coronavirus_2020}, and the Covid-19 Stringency Index (CSI) from the Oxford Coronavirus Government Response Tracker (OxCGRT) developed by \cite{hale_global_2021} (now referred to as CSI). The CSI describes the stringency of government measures by recording the number of government policies in each country and their strictness. The index is a composite measure based on nine response indicators, including school and workplace closures, travel bans, etc. These indicators are rescaled to a value from 0 to 100 (100 = strictest response). 
Together, these three datasets represent the health and social representation of the effects of Covid-19 on each country. The CSI has informed studies in the health sciences, such as estimating the impact of various physical distancing measures on disease incidence \citep{edejer_projected_2020} and relating different levels of healthcare resources to the associated transmission risk \citep{islam_physical_2020, hale_government_2021}. Political scientists have employed the CSI to consider whether stringency measures vary by regime type \citep{hale_pandemic_2020, frey_democracy_2020}, and whether upcoming elections influenced the strength of responses \citep{pulejo_electoral_2020}.

We source the datasets from the \textit{Our World in Data} `Data Explorer', which formats and aggregates a variety of datasets from academic and public institutions globally \citep{ritchie_coronavirus_2020}. \textit{Our World in Data} sources data on worldwide Covid-19 cases and deaths from the Covid-19 Data Repository by the Center for Systems Science and Engineering (CSSE) at Johns Hopkins University \citep{dong_interactive_2020}.

\begin{figure}[t]
\includegraphics[width = 0.93\textwidth]{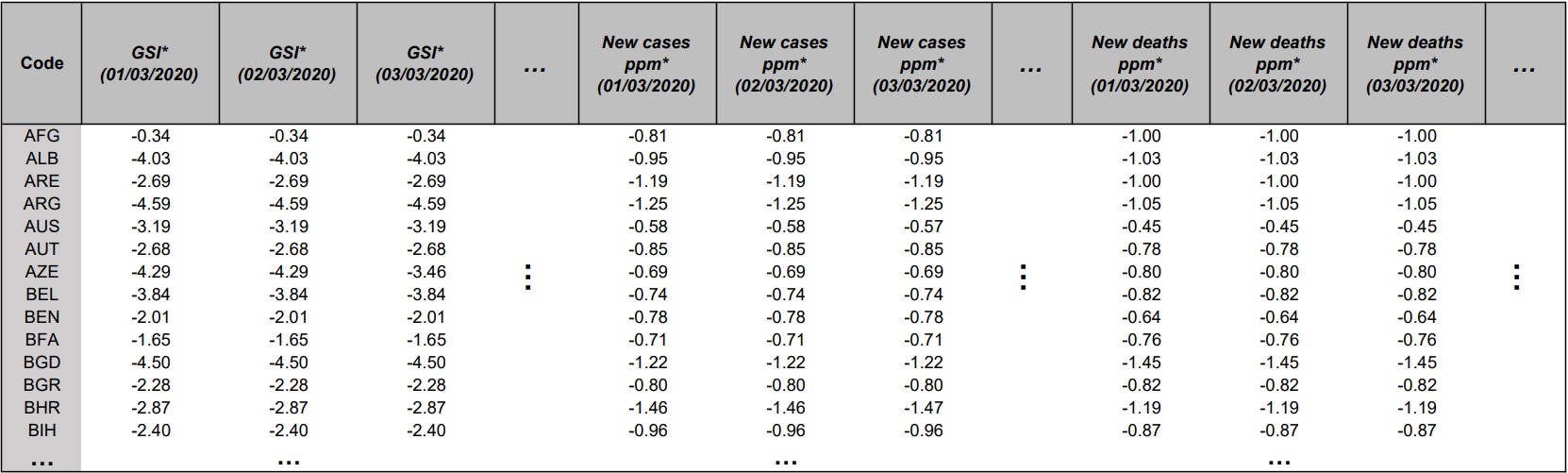}
\caption{{\bf Input dataset excerpt.}
Data format utilised in this analysis. The asterisk (*) denotes a dataset standardised to a z-score via Eq.~\eqref{zscore}. Here, pmp refers to `per million population'.}
\label{data_excerpt}
\end{figure}

We arrange the combined dataset with one row corresponding to a country, and the relevant time-series datasets of cases, deaths, and stringency index in columns. Fig. \ref{data_excerpt} provides an excerpt of the combined dataset.

The dataset spans a period of 454 days from $1^{\textrm{st}}$ Mar 2020 to $29^{\textrm{th}}$  May 2021 and initially included 190 countries. Given that this analysis includes three datasets each containing 454 temporal measurements, the nominal dimension of the dataset is $\textit{D} = 454 \times 3 = 1362$. 

To improve the robustness of the study, we included countries if they meet certain data availability and size requirements. We excluded countries with more than 20\% of missing data in the given period for any dataset. Any remaining missing values are imputed with a linear regression using the \texttt{imputeTS} package in \textsf{R} \citep{moritz_imputets_2017, R_core_team_2021}. Additionally, countries with smaller populations ($< 1$ million) can display higher volatility in new cases and deaths pmp. Thus, such countries were excluded from the dataset to limit the influence of outliers in the analysis. These data pre-processing procedures leave $n=115$ countries. 

All data manipulation or transformation tasks performed as part of the pre-processing methodology was undertaken using through the tools available in the \texttt{tidyverse} package in \textsf{R} \citep{tidyverse_ref, R_core_team_2021}. The completed pre-processed dataset is available in \nameref{S1_File}, and the corresponding code to replicate the data preparation methodology, results, and figures in this paper are available in \nameref{S2_File}. 

Additionally, the original dataset may be temporally stratified into four equally separate pandemic stages to reveal the ID at a more granular level in the dataset. The original dataset described in Table \ref{tbl:stage-dates} below documents the date range of each stage.

\begin{table}
\caption{\label{tbl:stage-dates}Date range for data subsets.}
\begin{tabular}{@{}c c c @{}}
\textbf{Stage} & \textbf{Beginning}  & \textbf{Ending}     \\ \hline
1              & $1^{\textrm{st}}$ Mar 2020    & $23^{\textrm{rd}}$   June 2020    \\
2              & $24^{\textrm{th}}$ June 2020    & $15^{\textrm{th}}$   October 2020 \\
3              & $16^{\textrm{th}}$ October 2020 & $6^{\textrm{th}}$  February 2021 \\
4              & $7^{\textrm{th}}$ February 2021 & $29^{\textrm{th}}$ May 2021 \\
\end{tabular}
\end{table}

We need to highlight some considerations before applying Hidalgo to this dataset. First, according to \citep{Allegra}, Hidalgo requires unique observations in a dataset and thus performs optimally on datasets with continuous data or discrete numbers within a broad range. Therefore, this analysis scales new cases and deaths by the country population to satisfy this assumption and enable disease dynamics to be compared across countries with different populations.  
Second, Hidalgo assumes identically and independently distributed observations in a dataset. This assumption cannot be fully satisfied in this application due to the inherent temporal autocorrelation present in the dynamics of cases, deaths, and corresponding government response measures. 
To limit temporal autocorrelation, two pre-processing steps are applied to the chosen datasets. Firstly, `new' cases, and deaths pmp are selected, as opposed to their `active' or `cumulative' counterparts. Additionally, each of the three datasets are normalised to z-scores across all countries, given by Eq.~\eqref{zscore} :

\begin{equation}
    z_{k} = \frac{x_{k} - \bar{x}_{k}}{s_{k}},
    \label{zscore}
\end{equation}
where $x$ represents a dataset, $k \in [1,2,3]$ denotes each of the three datasets used in this analysis, and $\bar{x}_{k}$ and $s_{k}$ represent the mean and standard deviation of a dataset respectively.

\subsection{Computational Details}
\label{sec:run}

Hidalgo was run on this dataset for 25,000 MCMC iterations ($nsim$), after a burn-in of 1,000 iterations. A sparse mixture modelling approach \citep{Rousseau2011,Malsiner-Walli2016} is employed in this analysis, with $L = 6$ mixture components, and $\alpha = 0.05$ for the Dirichlet priors of the mixture weights \citep{denti_intrinsic_2022}. Three matrices are produced as the output \citep{denti_intrinsic_2022}:
\begin{enumerate}
    \item Membership labels (dim: $nsim \times n$) where each column contains the MCMC sample of the membership labels for every observation;
    \item Cluster probabilities (dim: $nsim \times L$) where each column contains the MCMC
sample of the mixing weights for each mixture component;
    \item Intrinsic dimensions (dim: $nsim \times L$) where each column contains an
MCMC sample for every ID parameter estimated in each cluster.
\end{enumerate}

The MCMC chains produced by Hidalgo may exhibit label-switching issues, which prevents direct extraction of inference from the MCMC output. Due to label-switching, mixture components can be discarded, emptied, or repopulated across iterations. 

To obtain a reliable clustering estimate, one can inspect the posterior co-clustering matrix $PCM=\{p_{ij}\}$ computed across the $n$ countries, where each entry $p_{ij}$ is defined as the proportion of times that countries $i$ and $j$ have been clustered together across the $nsim$ MCMC iterations. Once the $PCM$ is estimated, one can recover a clustering estimation by minimising a loss function over the space of the possible partitions. A widely used method is the minimisation of the Variance of Information (VI) loss function \citep{meila_comparing_2007}. In this way, we can estimate the number of latent ID manifolds in the dataset. Moreover, we can also obtain more specific results by following a post-processing procedure described by \cite{denti_intrinsic_2022}, devised to address the label-switching issue. Indeed, label-switching arises whenever a mixture model with a-priori symmetric components is adopted. The algorithm that is used maps the $L$ different parameter-specific chains -- one for each mixture component parameter $\{d_l\}_{l=1}^L$ -- to $n$ observation-specific chains $\{d_{c_i}\}_{i=1}^n$. This way, not only are we able to draw inferences about the clusters characterised by heterogeneous ID present in the data, but we can also focus on the observation-specific ID estimates. Thus, in our application, we can compare the different country-specific ID estimates in addition to ID estimates of latent manifolds in the dataset.

\section{Results \& Discussion}
\subsection{Global Covid-19 data is characterised by low complexity}
\begin{figure}
\includegraphics[width = 0.93\textwidth]{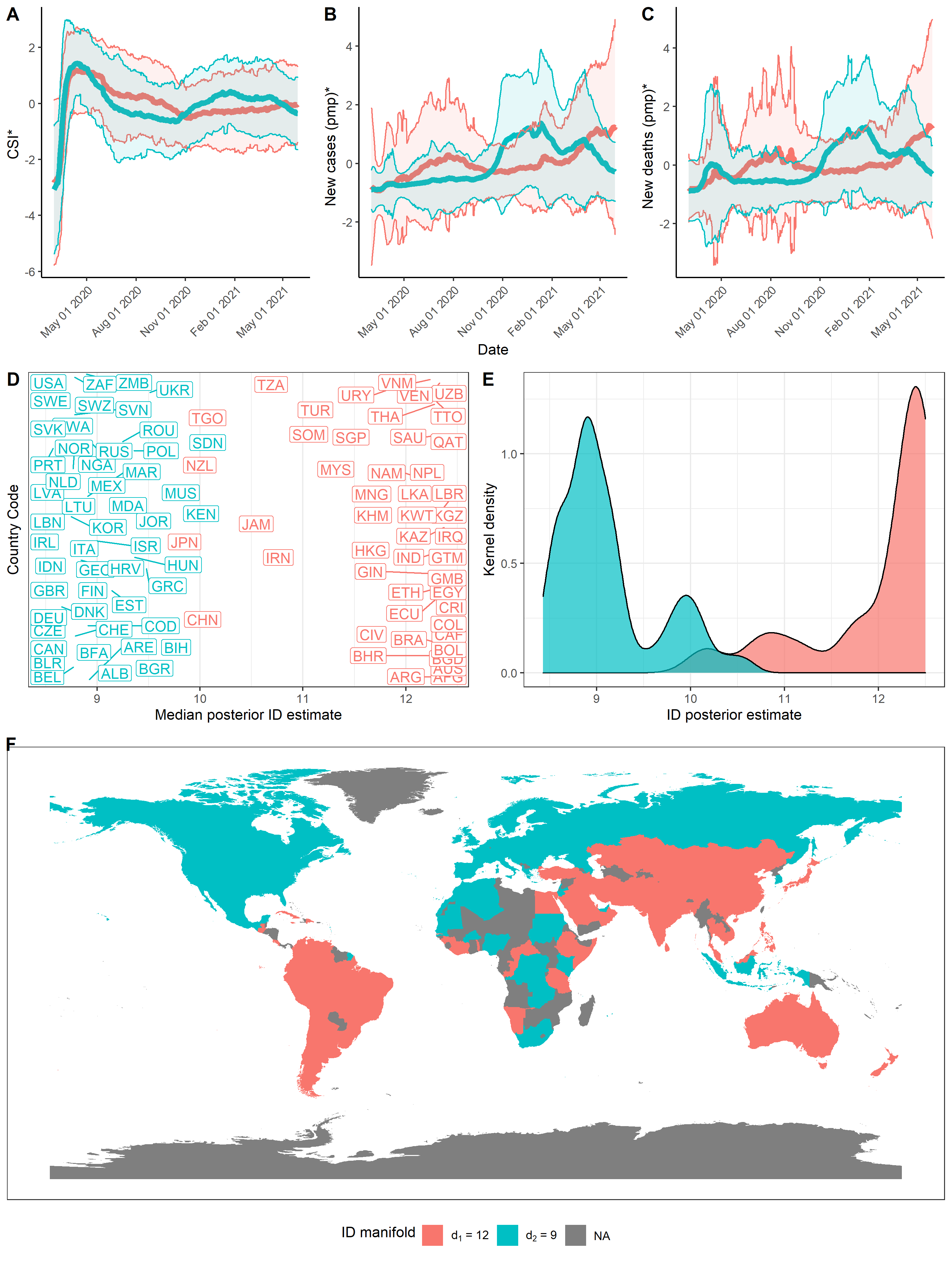}
\caption{{\bf Summary of results over the time period from $1^{\textrm{st}}$ Mar 2020 to $29^{\textrm{th}}$  May 2021.}
(A) Mean and standard deviation of standardised CSI aggregated by ID manifold, (B) standardised new cases pmp, and (C) standardised new deaths pmp. (D) Median posterior ID estimate by country, (E) posterior ID density estimated by manifold (E), and (F) world map of countries, coloured by ID manifold.}
\label{fig-full_time}
\end{figure}

Fig \ref{fig-full_time} presents a summary of the analysis. In particular, Fig \ref{fig-full_time}E highlights the posterior distribution of IDs in each cluster group, from which we can obtain a visual estimate of the variability of the ID estimates in each cluster. Hidalgo identifies two manifolds of posterior mean IDs $d_1=12$ and $d_2=9$, indicating the Covid-19 disease dynamics and corresponding government-established non-pharmaceutical interventions (NPIs) display higher redundancy in some countries than others. Countries assigned a higher ID indicate complex dynamics, as Hidalgo identifies these points project onto a high-dimensional manifold. Conversely, countries with a lower ID suggest simpler dynamics, as Hidalgo identifies these points project onto a low-dimensional manifold.

Given the high dimensionality of the dataset, IDs of 12 and 9 represent a dimensionality reduction of 99.34\% and 99.11\% respectively, suggesting strong dependence in the standardised new cases pmp, new deaths pmp, and the CSI for a country over the given period. Notably, these results indicate that a small set of parameters govern the Covid-19 dynamics, which has important implications for practitioners seeking to model these dynamics or apply dimensionality reduction techniques. For example, \cite{pope_intrinsic_2020} identify that lower IDs lower the sample complexity of learning, enabling more accessible learning for neural networks and better model generalisation from training to test data. 

Despite the overall low dimensionality of the dataset, the two ID manifolds identified differ by at least three dimensions. This result warrants further inspection to examine potential explanations for the dimensionality of each ID manifold.

\subsection{Global distribution of Covid-19 data complexity demonstrates spatial autocorrelation}
Notably, we identify evidence of spatial autocorrelation in the ID of global Covid-19 data, supported by Fig \ref{fig-full_time}D and \ref{fig-full_time}F. Fig. \ref{fig-full_time}D highlights the individual ID of each country included in the dataset. The colour of each country code corresponds to the ID manifold to which each country belongs. The ID manifold of each country may also be presented geographically on a map, as displayed in Fig. \ref{fig-full_time}F. Upon visual examination, Fig. \ref{fig-full_time}D and \ref{fig-full_time}F demonstrate that countries geographically close together tend to belong to the same ID manifold.

We confirm these results using Moran's \texttt{I} test, which is a widely used spatial statistic for detecting spatial autocorrelation \citep{moran_notes_1950, jackson_modified_2010}. Moran's \texttt{I} ranges from -1 to 1, and is defined as:

\begin{eqnarray}
\label{eq:moranI}
	{\displaystyle I={\frac {N}{W}}{\frac {\sum _{i}\sum _{j}w_{ij}(x_{i}-{\bar {x}})(x_{j}-{\bar {x}})}{\sum _{i}(x_{i}-{\bar {x}})^{2}}}}
\end{eqnarray}

where $N$ is the number of spatial units indexed by $i$ and $j$; $x$ is the individual median posterior ID estimate of a country; $\bar {x}$ is the mean of $x$ ; $w_{ij}$ is a neighbour adjacency matrix with zeroes on the diagonal (i.e., $w_{ii}=0$); and $W$ is the sum of all $w_{ij}$ \citep{moran_notes_1950}. In line with common approaches, we assign a weight of 1 for neighbouring zones and 0 otherwise \citep{jackson_modified_2010}. A neighbourhood is defined such that every country has at least one neighbour in the spatial weights matrix.

Applying Moran's \texttt{I} to the geographical distribution of median posteriors of ID produces an \texttt{I} value of 0.85 ($p<0.001$) using the \texttt{spdep} package in \textsf{R} \citep{spdep_ref, R_core_team_2021}, indicating significant positive spatial autocorrelation. 

This last result is a compelling finding, as the input dataset does not include any information about the geographical location of each country. Neighbouring countries may share the complex dynamics of Covid-19 as the pandemic spreads worldwide, resulting in positive spatial autocorrelation in the distribution of median posterior IDs of countries included in the analysis. \cite{mclafferty_placing_2010} suggests that geographically close countries are likely to share spatio-temporal dynamics due to human spatial dynamics and similar demographic factors across geographic regions. In reviewing the available literature, \cite{mcmahon_spatial_2022} highlight that a country's interconnectedness influences the spreading dynamics of Covid-19. This literature suggests that geographical closeness and interconnectivity have substantial implications for the spreading dynamics of Covid-19, allowing this to be a potential explanation for the spatial autocorrelation identified in the complexity of spreading dynamics observed in the analysis.
\subsection{Lower complexity data characterises high-income countries}
\begin{figure}
\includegraphics[width = 0.93\textwidth]{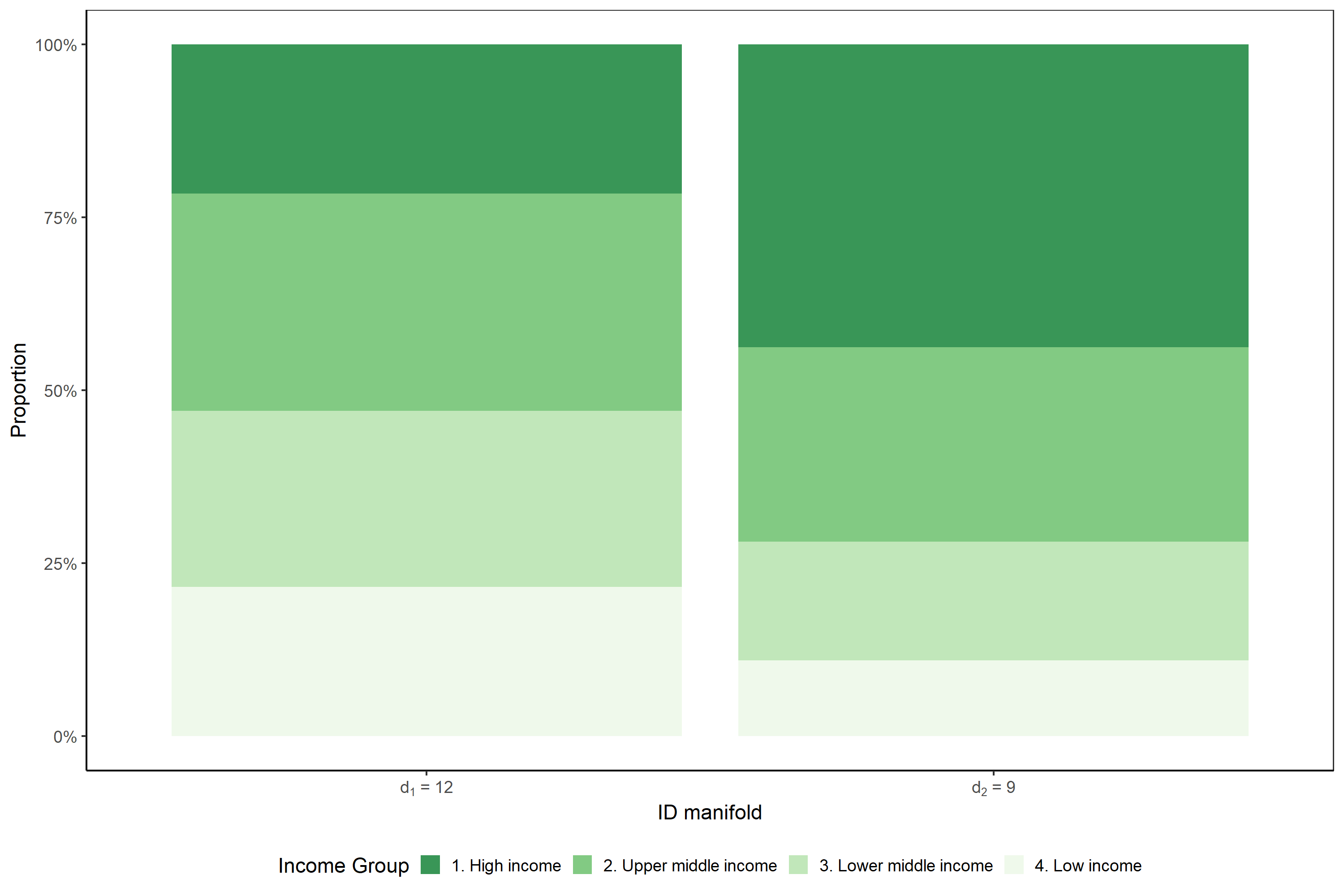}
\caption{{\bf Distribution of countries by income group in each ID manifold.} Income group classifications are retrieved from the World Bank database \cite{hamadeh_new_2021}.}
\label{fig-income_dist}
\end{figure}

\begin{figure}
\includegraphics[width = 0.93\textwidth]{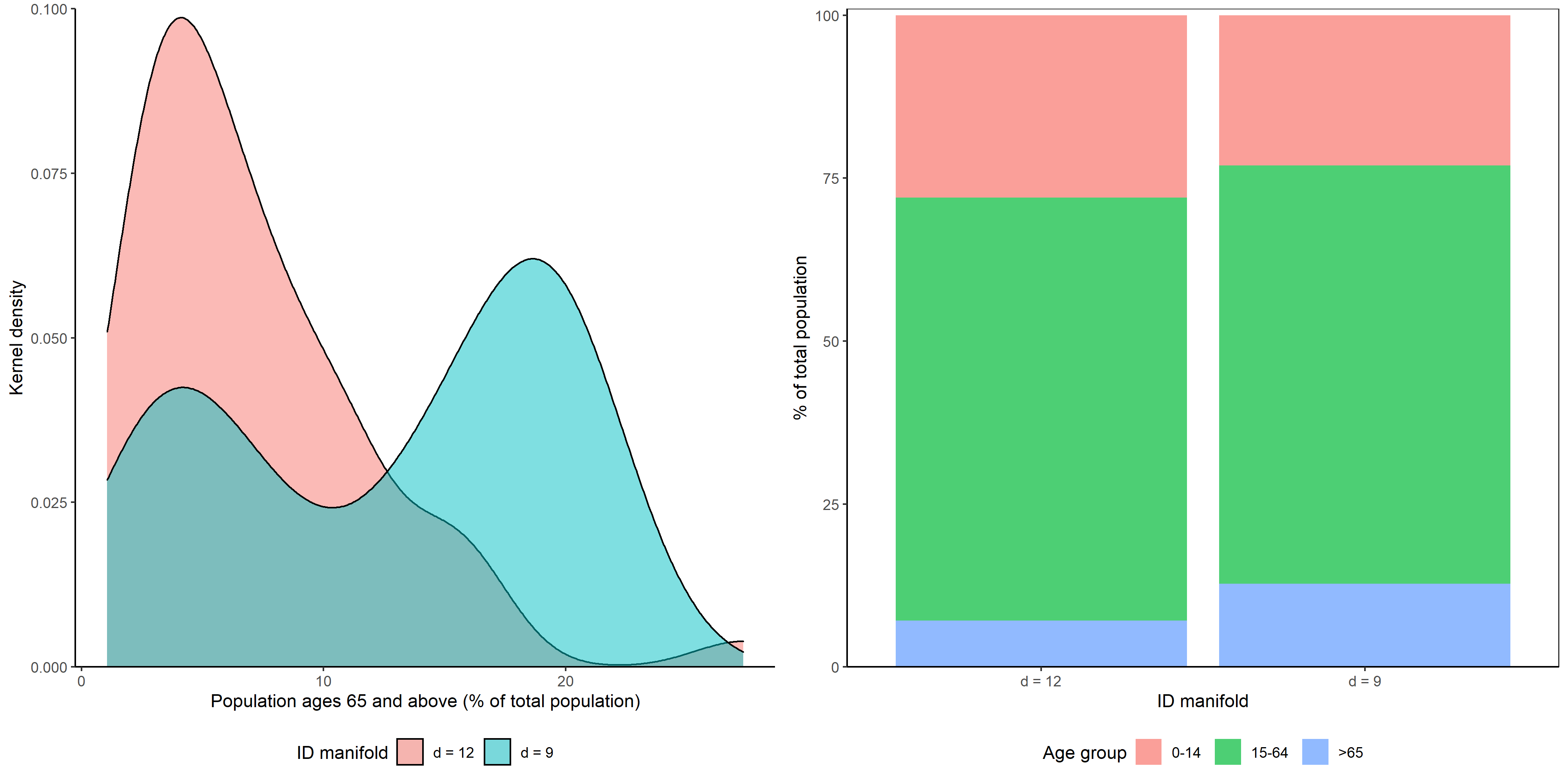}
\caption{{\bf Population distribution of countries in each ID manifold.}
A: Kernel density estimation of population ages 65 and above (\% of total), by ID manifold. B: Mean population distribution of countries by ID manifold.}
\label{fig-age_dist}
\end{figure}

Our analysis reveals that countries with higher income level groups are more likely to lie in low-dimensional manifolds. Fig. \ref{fig-income_dist} presents the distribution of income levels across the two ID manifolds. The World Bank assigns one of four income levels to each country, ranging from low- to high-income \citep{hamadeh_new_2021}. For the 2022 fiscal year, low-income countries fall under a Gross National Income (GNI) per capita of \$1,045 (USD) or less in 2020; lower-middle-income between \$1,046 and \$4,095; upper-middle-income between \$4,096 and \$12,695; and high-income from \$12,696 or more. GNI per capita represents the value produced by each person in a country's economy in a given year, regardless of whether the source of the value created is domestic production or receipts from overseas. While the GNI per capita does not entirely summarise a country's level of development or measure welfare, it has proved to be a useful and readily available indicator that closely correlates with other, non-monetary measures of the quality of life, such as life expectancy at birth, mortality rates of children, and enrollment rates in school \cite{international_economics_department_per_1989}.
A range of factors may explain the skewed distribution of income levels towards the low-ID manifold. High-income countries usually have aging populations, arising from declining fertility and improving mortality due to income growth, changes in health behaviours, and higher education levels \citep{ho_recent_2018, lee_health_2018}. Aging populations in many high-income countries have played a role in creating a greater mortality burden during the Covid-19 pandemic over 2020 to 2021 due to increased vulnerability to serious infections in population groups aged over 65 \citep{schellekens_covid-19_2020}. Additionally, underlying diseases such as diabetes, cardiovascular disease, and other diseases significantly contribute to increased severity risk from Covid-19 \citep{booth_population_2021}. Importantly, \cite{ofori-asenso_recent_2019} highlight that such chronic medical conditions are widely prevalent in aging populations in high-income countries. These factors have significantly impacted the mortality burden per capita over the Covid-19 pandemic. Research from the World Bank (\citeyear{schellekens_covid-19_2020}) and \cite{bayati_why_2021} estimate that high-income countries have had 2 to 3 times the Covid-19 mortality burden per capita compared to other countries over 2020. The age distribution disparity across the two ID manifolds is evident in Fig \ref{fig-age_dist}A and \ref{fig-age_dist}B.
Fig \ref{fig-age_dist}A reveals that countries assigned to a high-ID manifold had less than 7\% of the population aged over 65 on average. In comparison, countries assigned to a high-ID manifold have 13\% of the population aged over 65 on average, despite displaying bimodality due to some low-income countries in the low-ID manifold. A Kolgomorov-Smirnov test may be applied to evaluate the null hypothesis that the distributions are sampled from a population with the same distribution, which is subsequently rejected at the $p < 0.001$ significance level \citep{noauthor_kolmogorovsmirnov_2008}. The mean age distribution of countries in each ID manifold presented in Fig \ref{fig-age_dist}B corroborates that countries assigned to a low-ID manifold host a higher proportion of the population aged over 65, while countries assigned to a high-ID manifold host a higher proportion of the population aged between 0 and 14.

A possible explanation for high-income countries being assigned a low-ID manifold arises after identifying a link between high-income countries, aging populations, and increased per capita mortality burden. Namely, since new deaths pmp are a subset of new cases pmp, increased new deaths pmp in high-income countries may provide explanatory power to new cases pmp, resulting in greater dependency between the two time-series datasets and thus requiring a lower ID. Conversely, lower rates of reported new deaths pmp in low-income countries would lower the dependence in the entire dataset included in the analysis requiring a higher number of dimensions to represent the data accurately.

Furthermore, issues in data quality in Covid-19 data in low- and middle-income countries are widely researched. They may be another factor contributing towards the existing distribution of countries and corresponding designations to ID manifolds \citep{schellekens_covid-19_2020}. \cite{whittaker_under-reporting_2021} identifies that under-reporting in deaths varies globally but is highest in low-income and fragile settings. Such data artifacts could lead to a higher number of unexplained values, thereby lowering the dependence in the dataset and requiring more dimensions to describe the data effectively.

\subsection{Changes in ID over stages of the Covid-19 pandemic.}
\begin{figure}
\includegraphics[width = 0.93\textwidth]{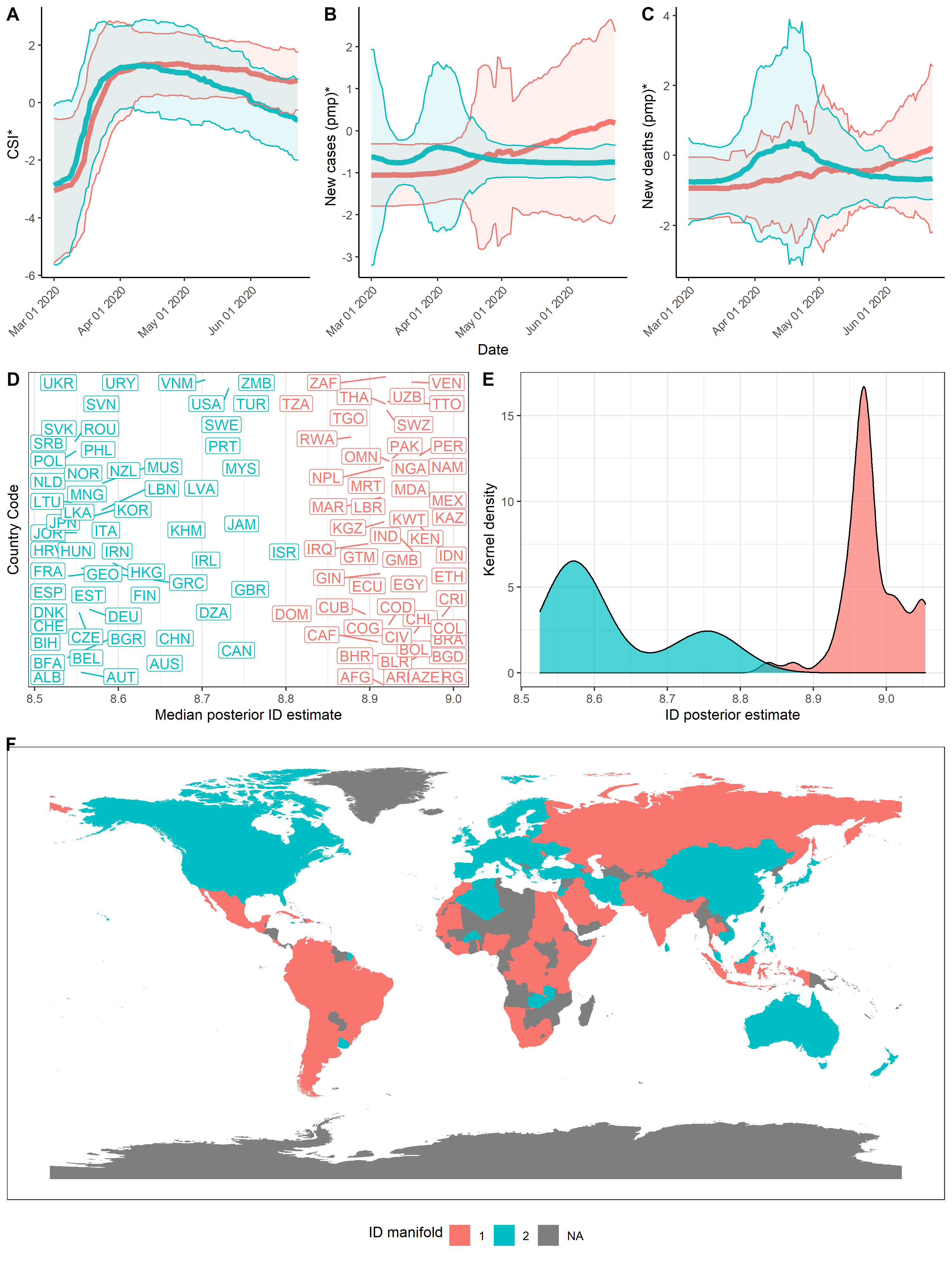}
\caption{{\bf Stage 1 - Summary of results over the time period from $1^{\textrm{st}}$ Mar 2020 to $23^{\textrm{rd}}$  June 2020.}
(A) Mean and standard deviation of standardised CSI aggregated by ID manifold, (B) standardised new cases pmp, and (C) standardised new deaths pmp. (D) Median posterior ID estimate by country, (E) posterior ID density estimated by manifold (E), and (F) world map of countries, coloured by ID manifold.}
\label{fig-stage_1}
\end{figure}

\begin{figure}
\includegraphics[width = 0.93\textwidth]{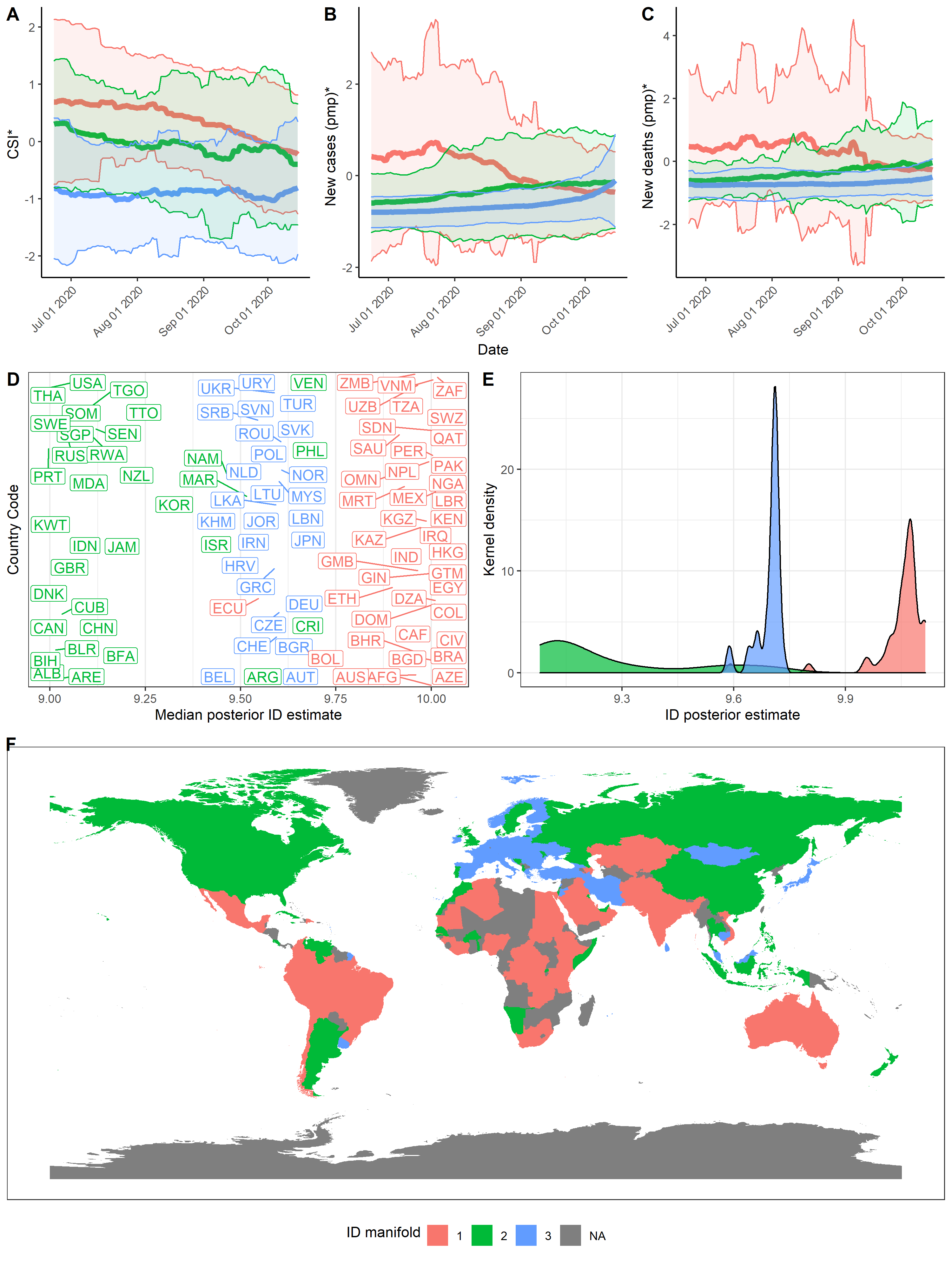}
\caption{{\bf Stage 2 - Summary of results over the time period from $24^{\textrm{th}}$ June 2020 to $15^{\textrm{th}}$ October 2020.}
(A) Mean and standard deviation of standardised CSI aggregated by ID manifold, (B) standardised new cases pmp, and (C) standardised new deaths pmp. (D) Median posterior ID estimate by country, (E) posterior ID density estimated by manifold (E), and (F) world map of countries, coloured by ID manifold.}
\label{fig-stage_2}
\end{figure}

\begin{figure}
\includegraphics[width = 0.93\textwidth]{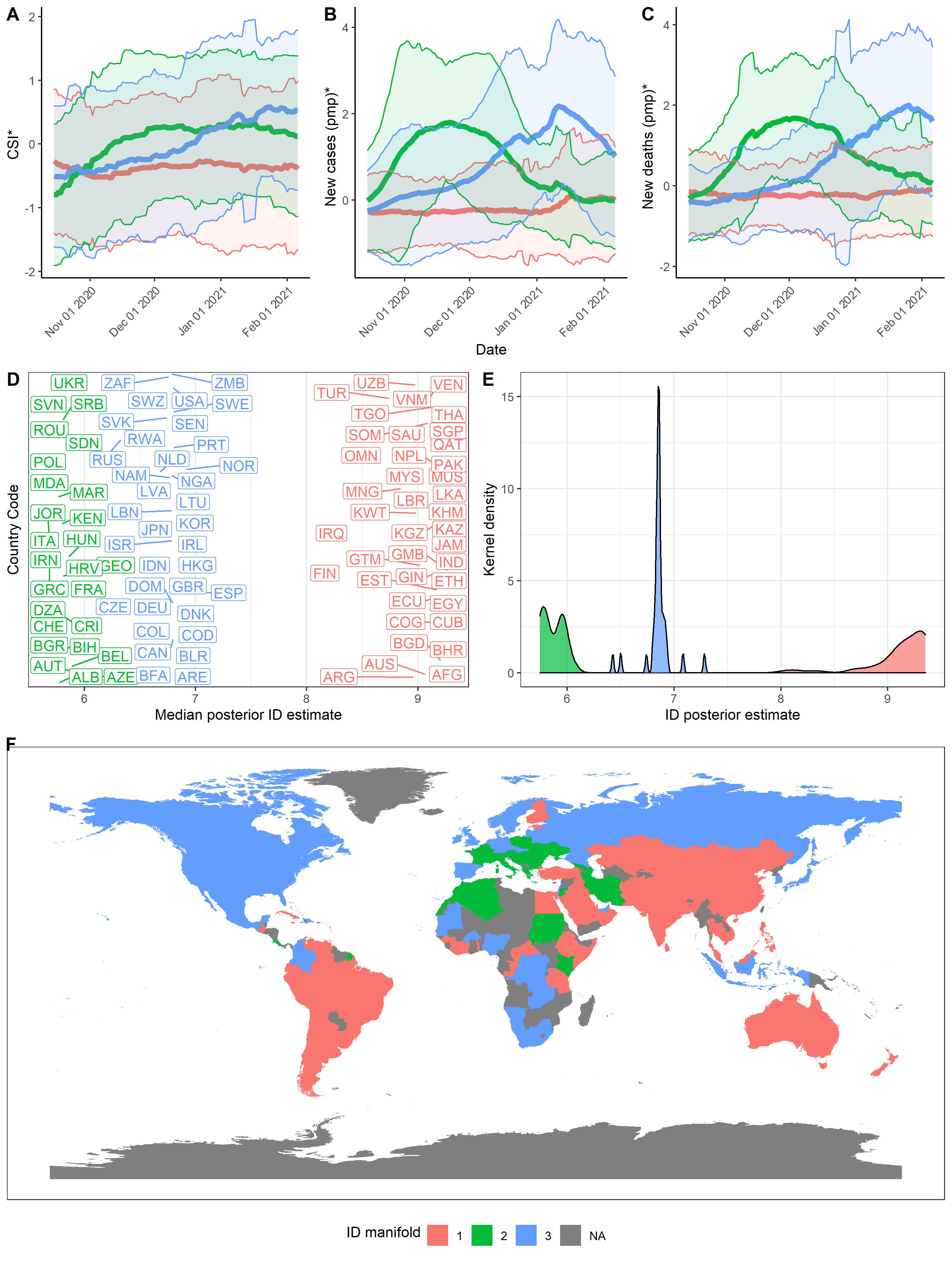}
\caption{{\bf Stage 3 - Summary of results over the time period from $16^{\textrm{th}}$ October 2020 to $6^{\textrm{th}}$ February 2021.}
(A) Mean and standard deviation of standardised CSI aggregated by ID manifold, (B) standardised new cases pmp, and (C) standardised new deaths pmp. (D) Median posterior ID estimate by country, (E) posterior ID density estimated by manifold (E), and (F) world map of countries, coloured by ID manifold.}
\label{fig-stage_3}
\end{figure}

\begin{figure}
\includegraphics[width = 0.93\textwidth]{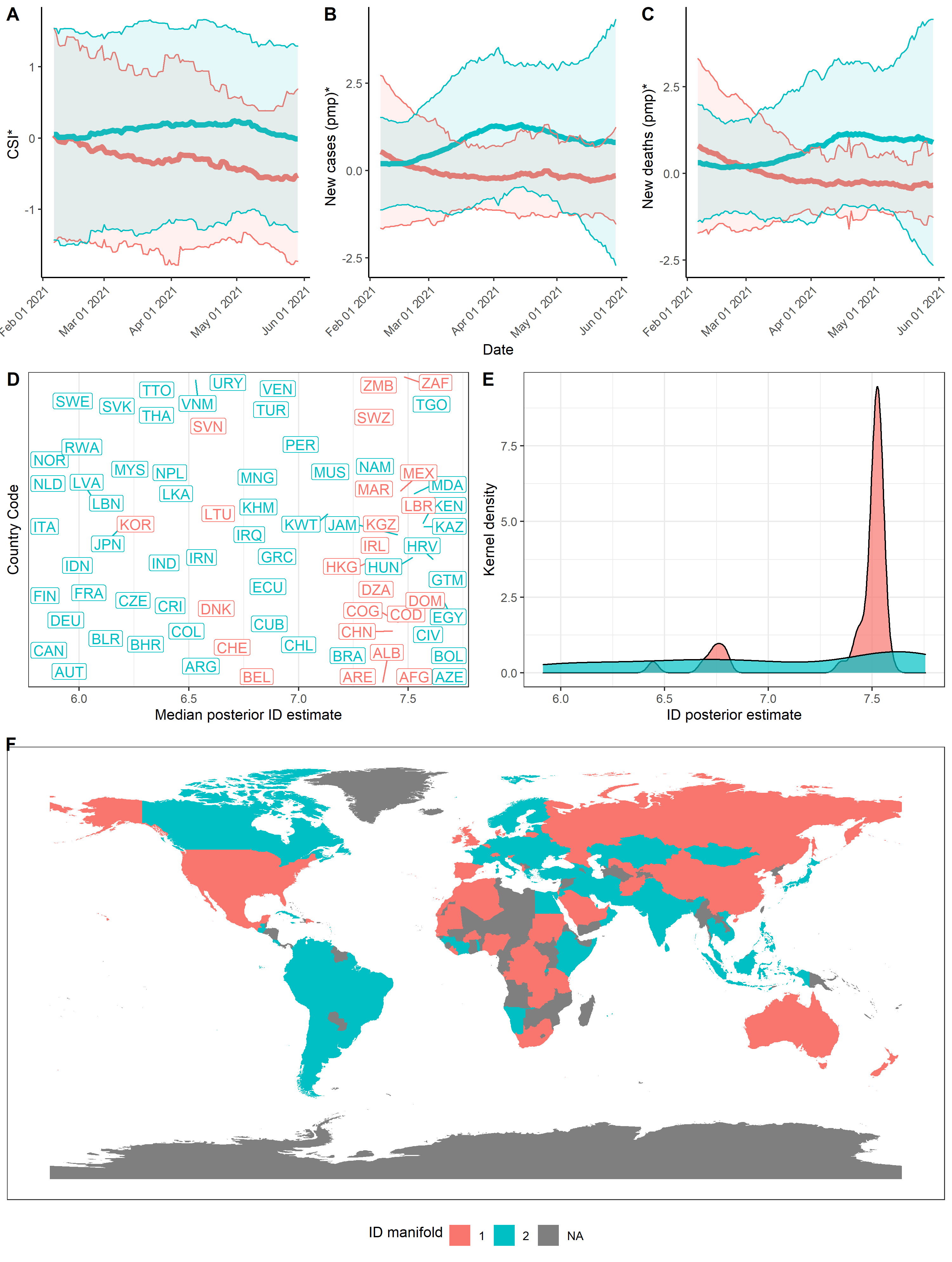}
\caption{{\bf Stage 4 - Summary of results over the time period from $7^{\textrm{th}}$ February 2021 to $29^{\textrm{th}}$ May 2021.}
(A) Mean and standard deviation of standardised CSI aggregated by ID manifold, (B) standardised new cases pmp, and (C) standardised new deaths pmp. (D) Median posterior ID estimate by country, (E) posterior ID density estimated by manifold (E), and (F) world map of countries, coloured by ID manifold.}
\label{fig-stage_4}
\end{figure}

Stratifying the datasets has provided a granular view of the ID over the course of the pandemic, and a summary of the results for each stage is presented in Figs. \ref{fig-stage_1}, \ref{fig-stage_2}, \ref{fig-stage_3} and \ref{fig-stage_4}. 

We can observe that countries lie between 2 to 3 ID manifolds throughout the pandemic. From March to June 2020 (Stage 1), manifolds have a similar ID which could reflect a generally united global response to the pandemic (Fig. \ref{fig-stage_1}, $d_1 = 9, d_2 = 8.6$). In June 2020 to October 2020 however, we find that the data lies on 3 different manifolds (Fig. \ref{fig-stage_2}, $d_1 = 10, d_2 = 9.2, d_3 = 9.75$). These 3 ID manifolds continue from October 2020 to February 2021, with all 3 manifolds lying on an ID between 9 and 10 (Fig. \ref{fig-stage_3}, $d_1 = 10, d_2 = 9.2, d_3 = 9.75$). Countries belonging to the manifold with the ID of 5.9 are mostly European (e.g., France, Italy). These countries experience a rise in the growth rate of cases and deaths, which precedes a corresponding rise in countries lying on the manifold with an ID of 6.9 (e.g., US, Spain, UK, Russia). Meanwhile, other countries with an ID of 9.2 continue to experience the average growth rates in cases and deaths (e.g., Australia, China, India, and much of South America). Finally, from February to May 2021, some countries lie on one clear manifold, with an ID of 7.5 (Fig. \ref{fig-stage_4}, $d_1 = 7.5, d_2 = 6.4$).

\subsection{Implications and future research}
We have successfully identified heterogenous ID manifolds for a dataset incorporating publicly available Covid-19 data such as government stringency levels, cases, and death rates per capita utilising Hidalgo, a Bayesian mixture model. Applying this model to the dataset reveals low intrinsic dimensionality, highlighting a potential for significant dimensionality reduction in the dataset. These findings suggest that few independent dimensions are required to effectively describe the dataset, enabling practitioners to discern better the level of model complexity required when describing or forecasting such data. 

Furthermore, we demonstrate how heterogenous ID estimators like Hidalgo may be employed to partition and simplify high-dimensional datasets. We reveal interesting spatial and demographic patterns in data that capture the unfolding of the global pandemic. It may be valuable for practitioners to consider these tools as part of their arsenal, to quantify data complexity and heterogeneity meaningfully,
as part of a quest to effectively extract useful information contained in high-dimensional data. 

Ultimately, the results of this analysis are subject to the quality of data available. While every effort has been made to correct for issues in the data, the inherent discrepancy in data quality across countries inevitably affects the results of this analysis. As previously ascertained by \cite{schellekens_covid-19_2020} and \cite{lloyd-sherlock_problems_2020}, it is currently infeasible to account for all under-reporting and data quality issues for specific countries and therefore remain an artifact in this dataset. It is also important to note that inherent assumptions in the Hidalgo algorithm require a careful choice of datasets in addition to some scaling transformations to limit temporal autocorrelation. These requirements limit the immediate applicability of Hidalgo for time series datasets as analysis must be conducted on standardised first-order differences with continuous values (e.g. new cases pmp, new deaths pmp), precluding practitioners from considering more intuitive datasets like cumulative cases or deaths.  

It would be valuable to conduct further examination on other factors contributing to the complexity (ID) of the Covid-19 data dynamics of a country to better understand drivers for complexity in pandemics. While we have identified that income level, age distribution, disease burden, and data quality all play a role in determining the ID of a country, developing a more nuanced understanding of these contributing factors would provide utility to the broader scientific community.

\newpage
\section{Conclusions}
This work evaluated the complexity of a dataset consisting of the standardised per-capita growth rate of Covid-19 cases, deaths, and an index describing a country's stringency of NPI measures (CSI), using a heterogenous intrinsic dimension estimator implemented as a Bayesian mixture model (Hidalgo). We identify that Covid-19 dataset may be projected onto two low-dimensional manifolds ($d_{1} = 12$, $d_{2} = 9$). Lower dimensionality suggests stronger dependence in the standardised growth rates of cases and deaths per capita and the CSI for a country over the given period. Notably, it indicates that Covid-19 data dynamics are governed by a small set of parameters, which has important implications for practitioners seeking to model these dynamics or apply dimensionality reduction techniques on this data. 

This work has demonstrated how the intrinsic dimension can help tell a story across multiple datasets and identify engaging ways to segregate data and retrieve novel insights. For example, we identify spatial autocorrelation in the distribution of ID estimates for countries. Furthermore, we highlight a skewed distribution of high-income countries projected on a low-dimensional ID manifold due to the increased per capita mortality burden from Covid-19 arising from aging populations and the increased prevalence of comorbidities. While we make significant progress towards understanding drivers for complexity in the included Covid-19 datasets, developing a more nuanced understanding of these contributing factors would enable decision-makers to better account for complexity in pandemics and is identified an area of future research.

\section{Supporting information}

\paragraph*{S1 File.}
\label{S1_File}
{\bf Covid-19 data utilised in this analysis.} This \texttt{.Rdata} file contains the pre-processed dataset developed to estimate the heterogenous ID manifolds with Hidalgo.

\paragraph*{S2 File.}
\label{S2_File}
{\bf \textsf{R} code to recreate the followed pre-processing pipeline, results and figures.} This \texttt{.R} file contains the code required to recreate the results and figures reported in this paper.

\section{Acknowledgments}
AV received funding from the Australian Research Council (ARC) Centre of Excellence for Mathematical and Statistical Frontiers for Big Data, Big Models and New Insights (ACEMS) under grant number CE140100049 and the First Byte Grant through the Centre for Data Science at the Queensland University of Technology. KM was supported by an ARC Laureate Fellowship under grant number FL150100150. AM was supported by FISR 2020 COVID No. FISR2020IP\_03843 and by European Union's Horizon 2020 research and innovation programme under grant agreement No. 101016233. The funders had no role in study design, data collection and analysis, decision to publish, or manuscript preparation.

\bibliography{references}
\end{document}